\documentclass[aps,prb,reprint,superscriptaddress]{revtex4-1}
\usepackage{graphicx}

\usepackage{color}

\begin{document}

%\preprint{}

\title{Non-adiabatic effects in the phonon dispersion of Mg$_{1-x}$Al$_x$B$_2$}


\author{Matteo d'Astuto}
\email[]{matteo.dastuto@impmc.upmc.fr}
%\thanks{}
%\affiliation{IMPMC, UMR CNRS 7590, Sorbonne Universit�s-UPMC University Paris 06, MNHN, IRD, 4 Place Jussieu, F-75005 Paris, France}
\affiliation{IMPMC, UMR CNRS 7590, Sorbonne Universit\'es-UPMC University Paris 06, MNHN, IRD, 4 Place Jussieu, F-75005 Paris, France}

\author{Rolf Heid}
\author{Burkhard Renker}
\author{Frank Weber}
%\affiliation{Institut f�r Festk�rperphysik, Karlsruher Institut f�r Technologie, P.O. 3640, D-76021 Karlsruhe, Germany}
\affiliation{Institut f\"ur Festk\"orperphysik, Karlsruher Institut f\"ur Technologie, P.O. 3640, D-76021 Karlsruhe, Germany}

\author{Helmut Schober}
\affiliation{Institut Laue-Langevin, BP 156 X, F-38042 Grenoble Cedex, France}

\author{Omar De la Pe\~na-Seaman}
%\affiliation{Instituto de F�sica, Benem�rita Universidad Aut�noma de Puebla, Apartado Postal J-48, C.P. 72570 Puebla, Puebla, M�xico}
\affiliation{Instituto de F\'isica, Benem\'erita Universidad Aut\'onoma de Puebla, Apartado Postal J-48, C.P. 72570 Puebla, Puebla, M\'exico}

\author{Janusz Karpinski}
\affiliation{Laboratory for Solid State Physics, ETH Z\"{u}rich,8093-Z\"{u}rich, Switzerland}%This line break forced% with \\ 

\author{Nikolai D. Zhigadlo}
\altaffiliation[Permanent address: ]{Department of Chemistry and Biochemistry, University of Berne, Freiestrasse 3, CH-3012 Berne, Switzerland }
\affiliation{Laboratory for Solid State Physics, ETH Z\"{u}rich,8093-Z\"{u}rich, Switzerland}%This line break forced% with \\ 

\author{Alexei Bossak}
\author{Michael Krisch}

\affiliation{European Synchrotron Radiation Facility, BP 220, F-38043
  Grenoble cedex, France}%This line break forced with \\


\date{\today}

\begin{abstract}
Superconducting MgB$_2$ shows an E$_{2g}$ zone center phonon, as measured by Raman spectroscopy, that is very broad in energy and temperature dependent. 
The Raman shift and lifetime show large differences with the values elsewhere in the Brillouin Zone measured by Inelastic X-ray Scattering (IXS), where its dispersion can be accounted for by standard harmonic phonon theory, adding only a moderate electron-phonon coupling. 
Here we show that the effects rapidly disappear when electron-phonon coupling is switched off by Al substitution on the Mg sites. Moreover, using IXS with very high wave-vector resolution in MgB$_2$, we can follow the dispersion connecting the Raman and the IXS signal, in agreement with a theory using only electron-phonon coupling but without strong anharmonic terms. 
The observation is important in order to understand the effects of electron-phonon coupling on zone center phonons modes in MgB$_2$, but also in all metals characterized by a small Fermi velocity in a particular direction, typical for layered compounds.  
\end{abstract}

%\keywords{}

\maketitle

%%\section{Introduction}

MgB$_2$ is a superconductor \cite{nagamatsu} where the mechanism for pairing is conventional electron-phonon coupling. It shows an unexpectedly high transition temperature of 39~K, almost twice that of most conventional systems at ambient pressure \cite{mazin-nat-views}, and close to those of high-temperature superconductors such as cuprates and iron pnictides. 
The phonon mode involved in the coupling is the transverse E$_{2g}$ mode, propagating along the c* direction with the atomic displacement parallel to the \textit{ab} plane (Ref. \onlinecite{bohnenPRL1,baron,mgb2-dastuto} and references therein).
Understanding the physics in MgB$_2$ is fundamental as it is the first and archetypal electron-phonon mediated superconductor with high-frequency phonons at ambient pressure \cite{floris-abini-mgb2-prl}. This is very important in the search of high temperature superconductivity, as these systems can be well modeled \cite{floris-abini-mgb2-prl,PhysRevLett.115.097002} and have potential for extremely high transition temperatures \cite{eremets-hs-nature}.

MgB$_2$ shows an anomalously low isotopic effect, surprising for a conventional pairing mechanism \cite{HinksNature}. It has been suggested that a strong anharmonic term, in the E$_{2g}$ mode energy, may explain this anomaly \cite{choi1}. The proposed strong anharmonicity seems in good agreement with the Raman data \cite{kunc,hlinka,quilty1,quilty2,bohnen,martinho} , but it is in contrast with IXS phonon dispersion results \cite{baron,mgb2-dastuto}, which are in good agreement with ab-initio models employing the quasi-harmonic approximation. However, a direct comparison is not possible as Raman spectroscopy only probes phonons close to the Brillouin zone center, $\Gamma$, where the IXS signal is dominated by elastic scattering.

In a previous work \cite{mgb2-dastuto}, it has been shown, from a close comparison of the temperature dependence of the IXS and Raman shift, that the apparent dichotomy between the two measurements could be reconciled taking into account non-adiabatic effects \cite{cappelluti} appearing only close to $\Gamma$. 
In that frameworks, anomalies in the Raman spectra have their origin in a very steep dispersion where the Landau-damping of phonons, induced by the electrons, disappears, as the phonon wave-vector is shorter than the Landau-damping threshold \cite{mgb2-dastuto}. This effect, which is very small and difficult to detect in most metals \cite{saitta,dean-na-gic}, would induce spectacular, temperature dependent, changes of the Raman width and shift in MgB$_2$.

Although the above explanation is quite reasonable, there is no complete proof of it, as the calculations are only qualitative, and we lack a direct observation of the suggested steep, anomalous dispersion close to $\Gamma$. 
Here we show, using high Q resolution Inelastic X-ray Scattering, that a very steep phonon dispersion can be detected in Mg$_{1-x}$Al$_x$B$_2$~ at very small x substitution, and rapidly disappears at higher Al content. 
The result is important for the theory of superconductivity of MgB$_2$, as an alternative interpretation of the Raman shift and broadening, in terms of anharmonicity, could explain the anomalous isotopic effect of MgB$_2$ \cite{yildirim}, but not the IXS shift and broadening \cite{mgb2-dastuto}. Moreover, there are very few experimental confirmations of non-adiabatic effects in solids, in particular their effect in the phonon dispersion, even if they can be relevant in the interpretation of vibrational data and the electron-phonon coupling of various layered metals such as intercalated graphite and graphene and, in general, for metals characterised by a small Fermi velocity in a particular direction in the Brillouin Zone \cite{saitta}. 

\begin{figure*}[t]
\includegraphics[width=0.49\textwidth]{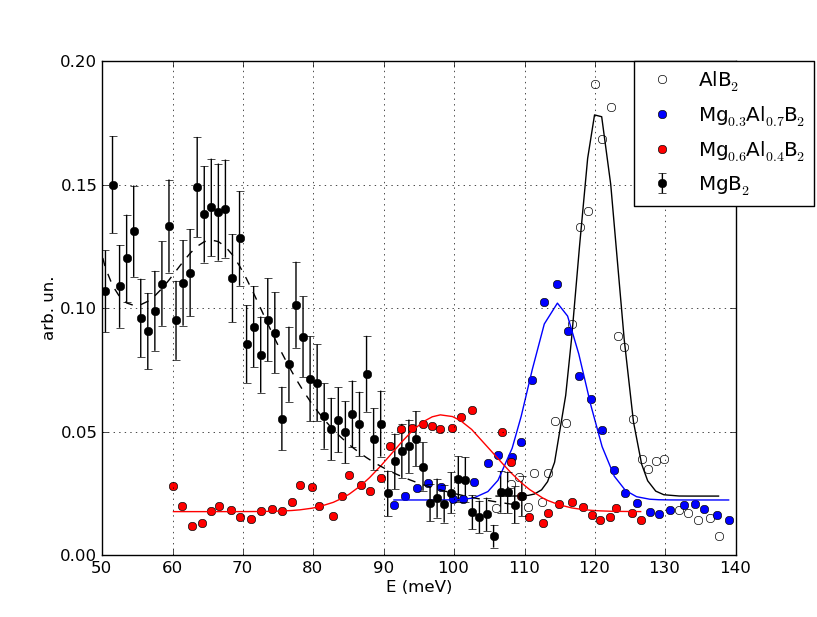}\includegraphics[width=0.49\textwidth]{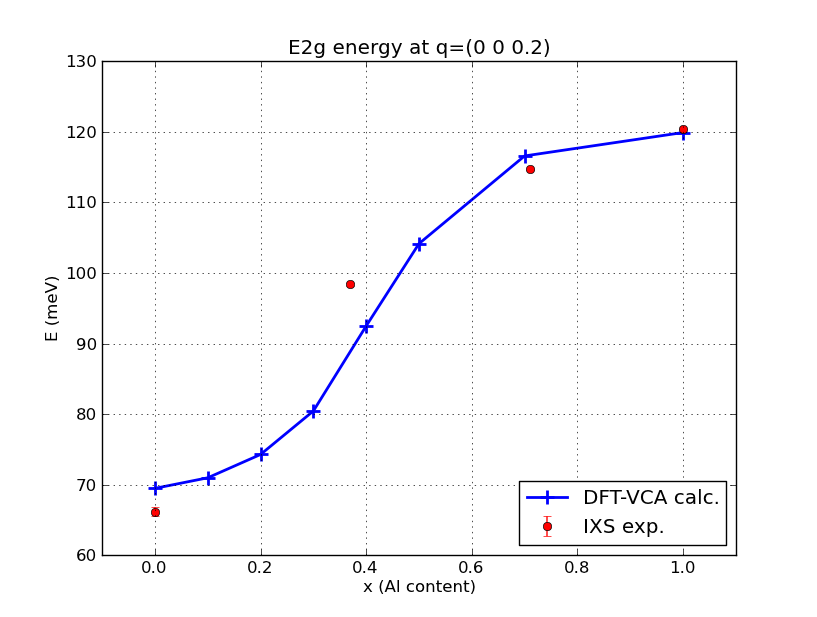} 
\caption{\label{compx}  (Color online) Left panel: IXS spectra (circle) and fit (lines) for Mg$_{1-x}$Al$_{x}$B$_2$ at the reduced wavevector q=(0 0 0.2), with colors coding the content x. Error bars can be inferred from the data dispersion. Lines are models of the data, with pseudo-Voigt fits (dashed, as in Ref.\onlinecite{mgb2-dastuto}) or simple Gaussian (continuous lines) profiles.
Right panel: Energy shift for the phonon at q=(0 0 0.2) as a function of the Al content x, from the fit of the data in the left panel (error bars from fit, mostly smaller than symbols), as well as from calculations in the Virtual Crystal Approximation (VCA).
}
\end{figure*}

%%\section{Sample preparation and experimental techniques}

For the crystal growth and characterization, see Supplemental Material at [URL will be inserted by APS]  as well as Ref. \onlinecite{mgb2-dastuto} for the MgB$_2$ samples. Note that in the text we rounded the precise stoichiometry given in Supplemental Material [URL will be inserted by APS] as follows for the Al content: 0.93 $\rightarrow$ 1 (note that there is no Mg in that sample); 0.71 $\rightarrow$ 0.7 and 0.37 $\rightarrow$ 0.4.

%%\subsection{Inelastic X-ray Scattering}

Inelastic X-ray Scattering (IXS) experiments were performed on the ID28 beam line of the European Synchrotron Radiation Facility.

The Mg$_{1-x}$Al$_x$B$_2$~ samples (0.37 $\leq$ x $\leq$ 1.0) were mounted in a vacuum chamber at room temperature such that the 110 and 001 directions were contained in the scattering plane. This orientation allowed for measurements of transverse and longitudinal branches along the $\Gamma$-A and $\Gamma$-M direction. 
For these samples, the back-scattering monochromator was set at the Si (8 8 8) reflection order \cite{verbeni,verbeni-rsi}, corresponding to a wavelength $\lambda$ = 0.7839 \AA ~ (photon energy: 15 817 eV) , and a resolution of $\Delta$E of 5.5 meV. 

IXS measurements in stoichiometric MgB$_2$~crystal were realized using higher energy resolution ( $\Delta$E = 3 meV) at the Si (9 9 9) reflection order of the back-scattering monochromator, in a very similar instrumental configuration of a previous experiment \cite{mgb2-dastuto} with the notable difference that we benefitted of 9 instead of five analyzers, allowing finer steps in the reciprocal space in particular for the \textit{in-plane} direction $\Gamma$-M, and taking particular care of adjusting the slit opening in front of the analyzers to maximize the spectrometer brilliance while optimizing the Q resolution as detailed in Supplemental Material at [URL will be inserted by APS]. 

\par
 
 All measurements were guided by calculations of inelastic structure factors resulting from first principles calculations. The spectra were analyzed by a fitting program which automatically accounted for the experimental resolution taken from measurements  or a model as described in Ref.  \onlinecite{mgb2-dastuto}. Typical scan times were about 2 hours.
Details of the ab-initio simulation in the Virtual Crystal Approximation (VCA) used in the present work are described in Ref. \onlinecite{Pena-Seaman_2009}. 

\begin{figure}[h]
\includegraphics[width=0.98\columnwidth]{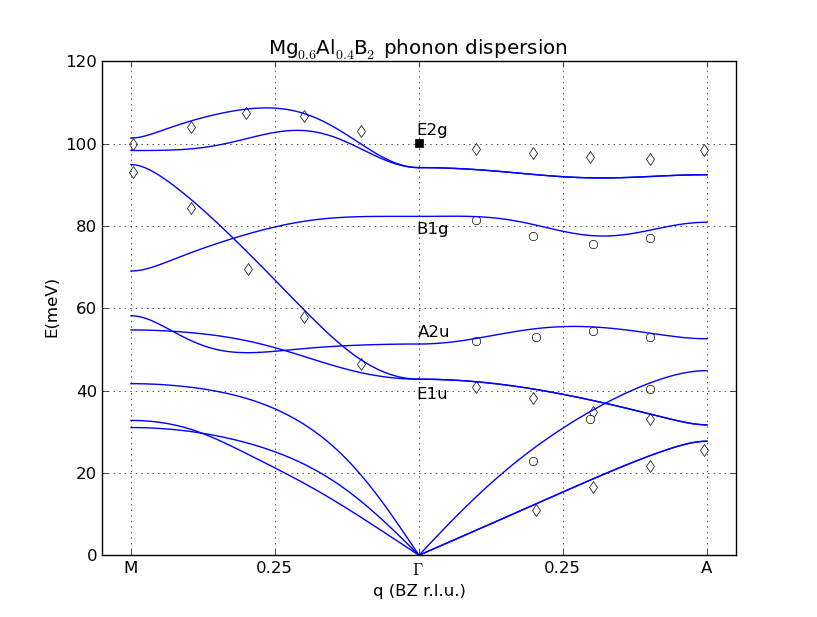}
\caption{\label{vca04} (Color online)   Calculated and measured dispersion of Mg$_{0.6}$Al$_{0.4}$B$_2$.
}
\end{figure}

%%  \section{Results}

Most of our data have been collected for the MgB$_2$ and Mg$_{0.6}$Al$_{0.4}$B$_2$ single crystals and the $\Gamma$-A direction which is of particular interest. 
In Fig. \ref{compx}, left panel, we show typical IXS spectra along the $\Gamma$-A line, at a reduced wave-vector \textbf{q}=(0 0 0.2), for several Al contents spanning from 0 to 1. in Fig. 
%\ref{shift_x} 
\ref{compx}, right panel, the energy of the E$_{2g}$ phonon as determined from the fit of these data is compared to the ones calculated for the same phonon using the Virtual Crystal Approximation approach (from Ref. \onlinecite{Pena-Seaman_2009} and present work for x=0.4). In Fig. \ref{vca04} we show a complete dispersion calculated for Al content x=0.4 along $\Gamma$-A and $\Gamma$-M high symmetry directions, in the VCA approach. 
The dispersion is compared with our IXS and Raman data for the sample with x=0.37 (Mg$_{0.6}$Al$_{0.4}$B$_2$).  

\begin{figure*}[t]
\includegraphics[width=0.49\textwidth]{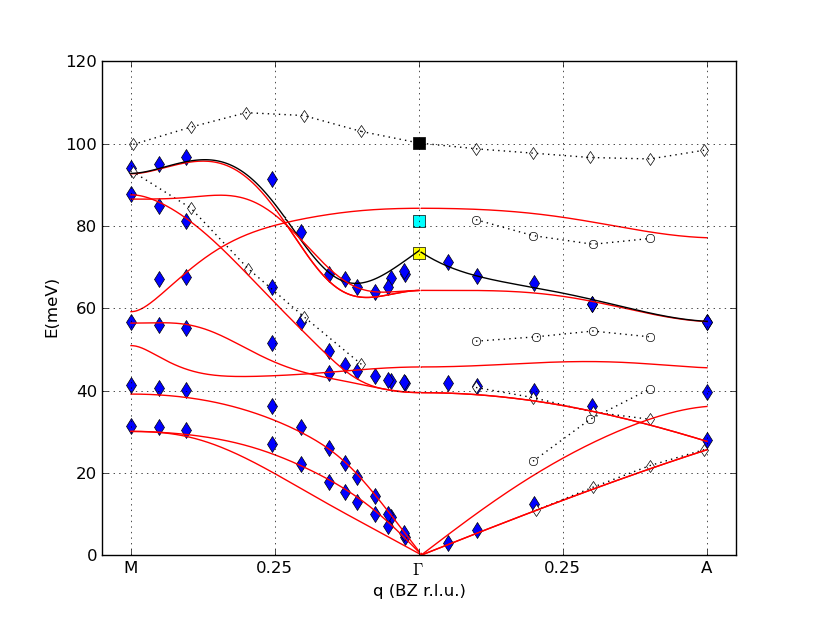}\includegraphics[width=0.49\textwidth]{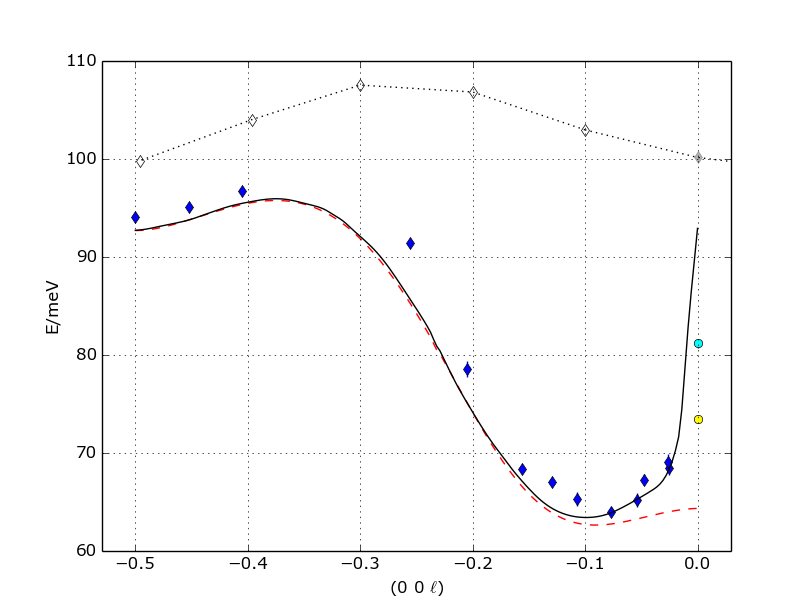}
\caption{\label{all_disp} (Color online) Left panel: Mg$_{1-x}$Al$_{x}$B$_2$ phonon dispersion along $\Gamma$-A and $\Gamma$-M directions: 
IXS data for x=0 (blue diamonds symbols) and 0.4 (white filled diamonds); Raman data for x=0 (filled squares, cyan, average values for T $\leq$ 100 K, and yellow, at ambient temperature, from Ref. \onlinecite{mgb2-dastuto}) and 0.4 (black square). 
Errorbars from the fit are smaller or of the same order as the symbol size. DFT calculation (red lines) for x=0 only (taken from Ref. \onlinecite{mgb2-dastuto}). The black line represents an empirical renormalization of the DFT calculation (see text).
Right panel: zoom on the E$_{2g}$ dispersion along the $\Gamma$-M direction. The dashed red line corresponds to the DFT calculation (red continous line in Fig. \ref{all_disp}). The black continuos line represent the "non-adiabatic" correction to the DFT calculations as explained in the text.
}
\end{figure*}

In Fig. \ref{all_disp}, left panel, we show the results of our present IXS measured dispersion for Al content x=0 (blue markers) and 0.4 (white filled markers) along $\Gamma$-A and $\Gamma$-M directions. We also add Raman data at x=0 (cyan and yellow filled symbols, from Ref. \onlinecite{mgb2-dastuto}) and 0.4 (black filled symbols). We compare with DFT calculations (red lines) for x=0 only, taken from Ref. \onlinecite{mgb2-dastuto}. The agreement is quite well overall, except for the the zone center Raman data at x=0. 
Calculation in  Ref. \onlinecite{mgb2-dastuto} and Ref. \onlinecite{Pena-Seaman_2009} are quite similar for x=0. An important difference, apart from the Virtual Crystal Approximation that allows Ref. \onlinecite{Pena-Seaman_2009} to change x, is that the authors of Ref. \onlinecite{Pena-Seaman_2009} use relaxed lattice parameters from the DFT calculation itself, while calculations in Ref. \onlinecite{mgb2-dastuto} use experimental lattice parameters, that give in general energies closer to the experimental ones, while relaxed lattice parameters tend to push some modes to higher energies, about 6 meV in the present case. This made the E$_{2g}$ mode energy calculated in Ref. \onlinecite{Pena-Seaman_2009} to match well the zone center energy but is about 6 meV higher at the zone boundary (point A).
Instead the E$_{2g}$ mode energy calculated in \cite{mgb2-dastuto} with experimental lattice parameters match well at the zone boundary (point A), as shown in Fig. \ref{all_disp}. To them, we added also an empirically renormalized dispersion (black lines) as guide to the eye of the zone center anomaly.

%%\section{Discussion}\label{discussion}
The overall evolution of the phonon modes along $\Gamma$-A and  $\Gamma$-M in function of x, going from x=0.0 to 0.4 qualitatively follows what it is expected according to Ref.  \onlinecite{Pena-Seaman_2009} for Al substitution in the x=0.3-0.5 range: a large hardening for the E$_{2g}$ mode ($\sim$ +35 meV) and a reduction for the A$_{2u}$ and longitudinal acoustic ones (few meV energy increase) with a concurrent moderate softening of the B$_{1g}$ mode (few meV decrease), with the E$_{1u}$ staying approximately at the same energy. 
A similar behavior is also experimentally observed for the E$_{2g}$ mode in carbon doped Mg(B$_{1-x}$C$_x$)$_2$ by IXS \cite{baron_physcaC_mgbc}.
Dispersion calculations for x=0.4 show a good quantitative agreement with our data; in particular we note that the frequency for the E$_{2g}$ mode at the zone center measured by Raman lies on an almost straight line  between the point at small $\mathbf{q}$ in the $\Gamma$-A and $\Gamma$-M directions, as in the VCA simulation.

As mentioned, this is not the case for x=0.0, for which there is a disagreement between the DFT calculations extrapolated to the zone center and the Raman data for the E$_{2g}$ mode, as can be appreciated in Fig. \ref{all_disp}, left panel. 
This was first ascribed to anharmonic effects \cite{choi1}, and seemed to be corroborated by the larger width of the Raman peak at room temperature, a factor two compared to the mode width measured by IXS along $\Gamma$-A, and with a large temperature effect that was not observed in IXS measurements \cite{baron}. 
In turn, IXS data seemed to fit quite well with the DFT calculation. 
Alternatively, it was later suggested that the effect could be ascribed to the crossing of the Landau damping threshold \cite{calandra}, meaning that, below a certain q wave-vector $q_0$ in the Brillouin Zone, the electron-phonon coupling cannot take place anymore, so that the phonon energy recovers from its value with electron-phonon coupling E to the bare one at q=0, E$_0$, which would be about 93 meV for the E$_{2g}$ mode. It follows that in the region 0$<q<q_0$, where the phonon modes decouple from the electrons, the system is not adiabatic anymore. This leads to a strong modification of the Raman spectral function, which probes the polarizability variation very close to (but not exactly at) $q=0$ as shown by Ref. \onlinecite{cappelluti}, with a larger broadening compared to the intrinsic one of the phonon modes, and a temperature dependence of such broadening in qualitative agreement with the measured one. 
The latter explanation also predicts an energy hardening with temperature, which was experimentally confirmed in
Ref. \onlinecite{mgb2-dastuto}. The reverse behavior would be expected from an anharmonic effect.

However, the proof was only qualitative, lacking a direct numerical comparison with data. 
The above mentioned recovery of E$_0$ would also induce a very steep dispersion as $q_0$ should be small (in the order of 0.1 of the BZ or less) while the energy shift is as large as $\Delta$ E = E-E$_0$ $\sim$35 meV as shown in Ref. \onlinecite{Calandra_PhysicaC}. Previous attempts to complete the dispersion close to the zone center $\Gamma$ were unsuccessful  (see Fig. 5 in Ref. \onlinecite{mgb2-dastuto}). In the present work we came back to perform measurements with a very high Q resolution and very fine step in Q close to $\Gamma$, from both the $\Gamma$-M and $\Gamma$-A directions. 
The data clearly bend up towards $\Gamma$ from both sides departing from the calculated dispersion. 
We stress again that this is not the case at all for the sample with x=0.4, where the Raman data falls exactly on a line going from $\Gamma$-M to $\Gamma$-A at small q, as expected from calculations at x$\sim$0.4 in Fig. \ref{vca04}. To highlight the effect on E$_{2g}$ at x=0, in Fig. \ref{all_disp} we modified the DFT calculated dispersion, multiplying it with an empirical exponential function $c*(1+e^{(-q/q_0)})$ (black lines, with $q_0\approx0.1$), as guide to the eye . We note here that the function is symmetric in $q$ units of the Brillouin Zone, but appears bending faster on the $\Gamma$-M side. This is due to the fact that the bare DFT calculated dispersion itself is already bending upwards close to $\Gamma$.   
We also note that a very steep phonon dispersion, known as the waterfall effect, was reported for relaxor perovskites materials \cite{PhysRevLett.84.5216} and explained in terms of mode-mode interference close to the zone center \cite{hlinka-wf}. However, in this scenario the two phonon modes involved have to have the same symmetry. As the only energy phonon branch in MgB$_2$ above the E$_{2g}$ has B$_{1g}$ symmetry, we can exclude a similar origin of the steep dispersion near $\Gamma$ in MgB$_2$. Moreover in Mg$_{0.6}$Al$_{0.4}$B$_2$ there is no steep dispersion at all, although the two modes are even closer in energy.

In order to see if our observations are compatible with crossing the Landau damping threshold, and the corresponding modifications of the dispersion, we compare with the calculations in Ref. \onlinecite{Calandra_PhysicaC}. In that work a modified dispersion along $\Gamma$-M is shown, including non-adiabatic effects mentioned above (see Fig. 3 left panel of Ref. \onlinecite{Calandra_PhysicaC}). But the E$_{2g}$ dispersion was a simplified, analytical one, and, although quite close to the E$_{2g}$ dispersion calculated by DFT, not accurate enough for a direct comparison with data, being \textit{e.g.} about 5 meV below the data for $q\sim~0.3-0.4~\times~\Gamma$-M, a shift about as large as the one we are looking for at $\Gamma$. This approximate dispersion was used because DFT can be performed neither exactly at the zone center, nor with enough q resolution to estimate the corrections of the self-energy for 0$<q<q_0$.
Conversely, the DFT calculated dispersion is much closer to the data, if we exclude the E$_{2g}$ mode in the $q$ region close to the zone center. 
In Fig. \ref{all_disp}, right panel, we zoom on the same E$_{2g}$ data as in the left panel, along $\Gamma$-M, with the red dashed line corresponding to the E$_{2g}$ dispersion calculated by DFT as in the left panel.
We extract then the correction due to a frequency dependence of the phonon self-energy from the dispersion in Ref. \onlinecite{Calandra_PhysicaC}, as the ratio between the "adiabatic" ($\epsilon^{adiabatic}_{\mathbf{q}E_{2g}}$) and "non-adiabatic" ($\epsilon_{\mathbf{q}E_{2g}}$) dispersion $\frac{\epsilon_{\mathbf{q}E_{2g}}}{\epsilon^{adiabatic}_{\mathbf{q}E_{2g}}}$. This is a rather crude approximation in principle, but the corrections are rather small in absolute values, so \textit{a posteriori} valid to first order. 

We apply the corrections to the DFT calculated dispersion (black line in Fig. \ref{all_disp}, right panel). This is justified by the fact that the analytic calculated $\epsilon^{adiabatic}_{\mathbf{q}E_{2g}}$ in Ref. \onlinecite{Calandra_PhysicaC} is already a rather good approximation of the real one. 
The result is astonishingly close to the data, hence supporting the view that the effect comes indeed from the inclusion of non-adiabatic effects, by computing the phonon self energy at finite phonon frequency when including the electron-phonon coupling as explained in Ref. \onlinecite{Calandra_PhysicaC}. 
Note that, while the zero temperature calculation recovers the unperturbed frequency (without electron-phonon coupling) at the zone center, finite temperature, and resolution, shift the Raman energy to lower values, with a positive temperature dependence that can not be explained by anharmonic effects as first pointed out in Ref. \onlinecite{cappelluti,mgb2-dastuto}.

%%\section{Conclusions}

In conclusion, we experimentally reconciled the IXS dispersion for the E$_{2g}$ mode in MgB$_2$ to the Raman data at the zone center, showing that the apparent difference can be explained without invoking any experimental artifact, but is connected to a very steep dispersion, as foreseen by theory based on anomalous effects of the electron-phonon coupling near the zone center \cite{cappelluti, Calandra_PhysicaC, saitta}, and does not require any strong contribution from phonon-phonon scattering (anharmonicity). 
This zone center anomaly disappears when the electron density-of-states at the Fermi level is depleted, by Al substitution x$\sim$0.4, concurrently with the loss of superconducting properties \cite{kortus-prl-c-al-mgb2}, a further indication that this anomaly arises from electron-phonon coupling effects. 

The observation is important to understand the effects of electron-phonon coupling on zone center phonons modes in MgB$_2$, and its relevance to its superconducting properties. 
Moreover, this is relevant in all metals characterized by a small Fermi velocity in a particular direction, and for which the temperature dependence of the Raman shift is often measured in order to detect electron-phonon coupling effects \cite{saitta}. 

\begin{acknowledgments}
We wish to acknowledge the ESRF for the support under his the non-propretary research program.
We are very grateful to D. Gambetti for technical help during IXS measurements. 
We acknowledge very useful discussion with F. Mauri, M. Calandra and M. Lazzeri, and we thanks J.
Hlinka for detailed clarification about the waterfall effect.  
F.W. was supported by the young investigator group VH-NG-840 of the Helmholtz Society.
\end{acknowledgments}

\bibliography{mgalb2}

\end{document}